\documentclass[pre,twocolumn,showpacs,amsmath]{revtex4}
\usepackage{graphicx,amsfonts,amsmath}

\newcommand{\bx}{\mathbf{x}}
\newcommand{\bn}{\mathbf{n}}

\newcommand{\bt}{\mathbf{t}}

\begin{document}

\title{Microscopic origin of granular ratcheting}
\author{S. McNamara}
\affiliation{Institut f\"ur Computerphysik, Universit\"at Stuttgart,
70569 Stuttgart, GERMANY}

\author{R. Garc\'{\i}a-Rojo}
\affiliation{Institut f\"ur Computerphysik, Universit\"at Stuttgart,
70569 Stuttgart, GERMANY}

\author{H.J. Herrmann}
\affiliation{Computational Physics, IfB, HIF E12, ETH H\"onggerberg CH-8093,
Z\"urich, SWITZERLAND}

\date{\today}

\begin{abstract}
Numerical simulations of assemblies of grains under cyclic
loading exhibit ``granular ratcheting'': 
a small net deformation occurs with each cycle, leading to a linear
accumulation of deformation with cycle number.  We show that this is
due to a curious property of the most frequently used models of
the particle-particle interaction: namely, that the potential
energy stored in contacts is path-dependent.  There exist closed
paths that change the stored energy, even if the particles remain
in contact and do not slide.  An alternative method for calculating
the tangential force removes granular ratcheting.
\end{abstract}

\maketitle

\section{Introduction}

Granular ratcheting refers to the slow, linear accumulation of strain
in a granular sample under cyclic loading.  Several versions of this
phenomena have been identified.  The first variant to be found occurs
when the loaded sample reaches the critical state once per cycle.
The mechanism is easily understood: the material flows while it is
in the critical state, giving rise to a deformation that accumulates
with cycle number.  However, ratcheting can also appear even when
the sample never reaches the critical state \cite{Festag}.  In the 
following, we discuss exclusively this second type of ratcheting.

Ratcheting in the absence of a critical state has also been observed
in numerical simulations \cite{Fernando,Ramon,Ramon2,RamonItaly}.  
This is a very promising development, for one has access to all the
quantities in numerical simulations, and it is usually possible to
identify the origin of the phenomena.  Once this has been done, one
can then ask if the cause of the phenomena
in the simulations is related to the cause in the experiments.

Numerical studies have already provided
many clues to granular ratcheting.  
The important role of sliding contacts has been pointed out \cite{Fernando},
granular ratcheting has been delimited from other possible behaviors
\cite{Ramon}, and the influence of various parameters has been studied
\cite{Ramon2,RamonItaly}.
One finding of these studies 
is that granular ratcheting is a quasi-static phenomena.  Specifically,
if one lets the frequency of the oscillating force tend to zero
while keeping all other parameters the same, the deformation per
cycle approaches a constant.

What is still missing is an understanding of granular ratcheting on
the micro-mechanical level:  
How exactly does the phenomenon arise from
interaction of individual particles?  Is it possible to modify the
particle interaction law to eliminate ratcheting?
What is the simplest system needed to produce ratcheting?  We answer
these questions in this paper.

In Sec.~\ref{ratchet16} we show that ratcheting can be obtained with
only 16 particles.  Ratcheting occurs if a single contact becomes
sliding.  Previously, ratcheting was linked to the presence of sliding
contacts, but this is the first time that it is shown that only a single
contact is necessary.  We also show that the application of
unconventional boundary conditions can lead to ratcheting even when
there are no
sliding contacts.  In Sec.~\ref{origin}, we show how granular ratcheting
arises from the way tangential forces are calculated.  In Sec.~\ref{sec:AMD}
we present an alternative method that does not exhibit ratcheting.
Finally, in the appendix, we show how stiffness matrix theory
can illuminate some aspects of the problem.

\section{Description of granular ratcheting}
\label{ratchet16}

\subsection{Model definition}
\label{model}

In this section, we present a very brief description of granular
ratcheting, since more complete discussions already exist
\cite{Ramon2}.  Granular ratcheting is observed in biaxial or
triaxial tests, where a granular sample is enclosed in a test
chamber, and subjected to a uniform pressure and a cyclic load.
We consider here exclusively the two dimensional version of these
experiments, often called the ``biaxial box'',
where a granular sample composed of disks is enclosed in
a rectangular box of dimensions $L_x \times L_y$, with forces 
$F_x$ and $F_y$ exerted on the walls.  The forces are
\begin{equation}
F_x = P_0 L_y, \quad F_y = L_x \left[P_0 + q(t)\right],
\label{imposed}
\end{equation}
where $P_0$ is the pressure exerted on the sample, and $q(t)$
is a periodic function, usually sinusoidal.
In the simulations presented here,
$q(t) = \Delta\sigma(1-\cos\omega t)$.
One usually uses deviatoric strain
\begin{equation}
\gamma = \frac{L_y}{L_{y0}}-\frac{L_x}{L_{x0}},
\label{defstrain}
\end{equation}
to characterize the deformation.
(Here $L_{x0}$ and $L_{y0}$ are the lengths of the system at the
beginning of the simulation.)  

We use a common numerical model of granular materials: grains are
represented by disks who repel each other when they overlap.  Thus
whenever two disks touch each other, they exert a repulsive force $F_n$
at the point of contact, directed normal to the particle surfaces.
$F_n$ is an increasing function of the overlap $D_n$.
If the surfaces of two touching disks move relative to each other
in the tangential direction,
a second force $F_t$ arises, directed tangent to the particle surfaces.
$F_n$ and $F_t$ are
called the normal and tangential components of the contact force.
In addition to these forces, some damping forces are included
to remove energy injected by the loading.

The contact force is subjected to two constraints, namely
\begin{equation}
F_n \ge 0, \quad \mu F_n \ge |F_t|.
\label{contactconditions}
\end{equation}
The first condition excludes cohesion, and the second is the
Coulomb friction law.  The constant $\mu$ is the Coulomb friction
coefficient.  Contacts where $|F_t|=\mu F_n$ are called sliding
contacts, and those where $|F_t| < \mu F_n$ are called
non-sliding.

All studies of granular ratcheting use this model, except sometimes
polygons are used instead of disks \cite{Fernando}.  

\subsection{Ratcheting with sixteen particles}
\label{sec:ratchet16}

If one wishes to approximate the continuum-like behavior of soils,
simulations with large numbers of particles are necessary.  Therefore,
granular ratcheting has been studied in assemblies of hundreds or
thousands of particles.  In this paper, however, we wish simply to
discover the origin of the phenomena, so it is useful to consider
small numbers of particles.  In this section, we study an assembly of
sixteen particles that exhibits granular ratcheting.  The normal
force is taken to be proportional to the overlap area, as in
Ref.~\cite{Ramon2}.

\begin{figure}[tb]
\centering
\includegraphics[height=0.5\textwidth,angle=-90]{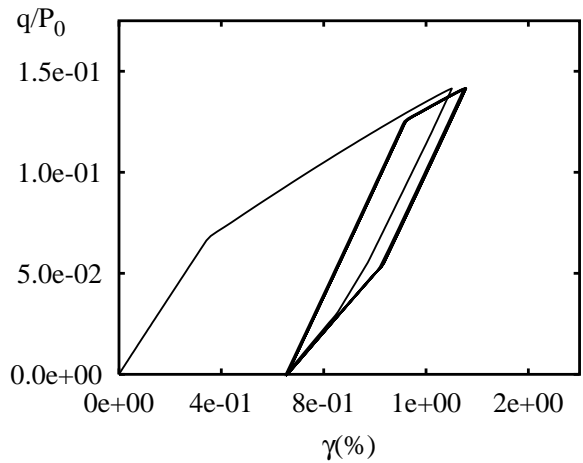}
\caption{Stress-strain curve of the system with $16$ particles discussed in
Sec.~\ref{sec:ratchet16}.}
\label{stress-strain}
\end{figure}

\begin{figure}[tb]
\centering
\includegraphics[height=0.5\textwidth,angle=-90]{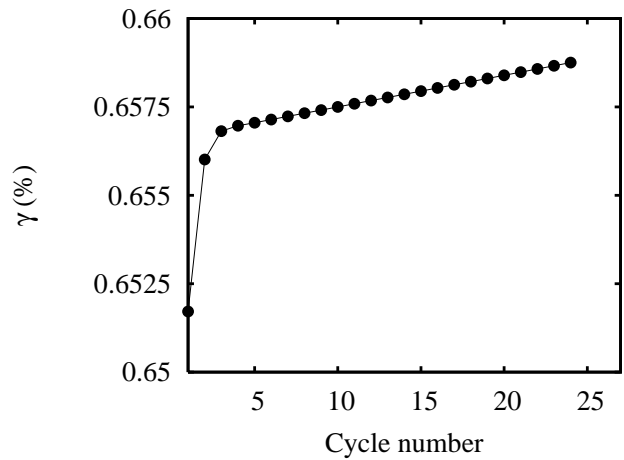}
\caption{Accumulated strain for the system shown in Fig.~\ref{stress-strain}.}
\label{accum-strain}
\end{figure}

In Fig.~\ref{stress-strain} we show a plot of $\gamma$ vs $q$ for a
biaxial test performed on sixteen circular particles.  In the first cycle,
$\gamma$ increases to about $0.008$.  During subsequent cycles,
the system appears to trace out a four-sided polygon in the $q$-$\gamma$
plane.  However, the path is not quite a polygon, because the system
does not quite return to its starting point after one cycle, but
to one where $\gamma$ is slightly larger.  This is made obvious
in Fig.~\ref{accum-strain}, where
$\gamma$ is plotted at $t=nT$, were $T$ is the period of the
cyclic loading, and $n=0,1,2\ldots$.
A small, linear increase of $\gamma$ with
cycle number is visible.  This is granular ratcheting.

During one cycle, all the contacts remain non-sliding, except for one, which
becomes sliding twice per cycle.  This single contact is responsible for
granular ratcheting, for if we inhibit sliding at this contact by increasing
$\mu$, granular ratcheting stops.  However, we show below that 
ratcheting can also occur without sliding contacts with slightly different
boundary conditions.

\begin{figure}[tb]
\centering
\includegraphics[height=0.5\textwidth,angle=-90]{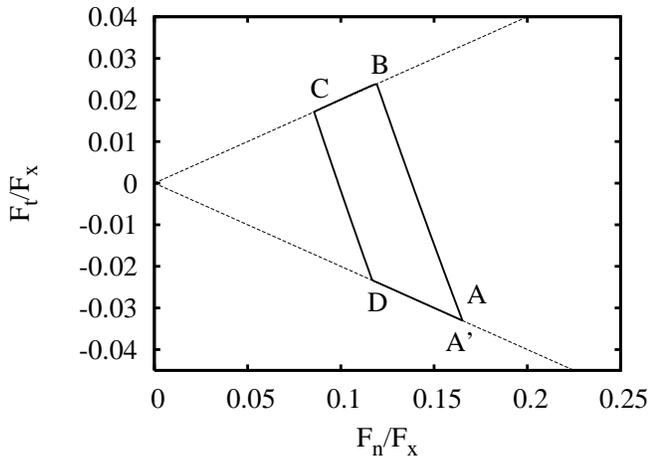}
\caption{A sketch of the sliding contact's trajectory in its
$(F_n,F_t)$ plane.  The diagonal dotted lines show the Coulomb
condition given in Eq.~(\ref{contactconditions}).}
\label{Fplane}
\end{figure}

In Fig.~\ref{Fplane}, we show the trajectory of the sliding contact in its
$(F_n,F_t)$ plane.   The equalities
$|F_t|=\mu F_n$ are also shown on the graph, and form a cone, with
the vertex at the origin.  The conditions in Eq.~(\ref{contactconditions})
mean that $(F_n,F_t)$ must always lie within this
cone.  As one can see, the ratcheting contact's trajectory is
a trapezoid, with the four corners
labeled $A$, $B$, $C$, and $D$.
The two parallel line segments correspond to the change
in force when all contacts are non-sliding.  Line segments $BC$ and $DA$
lie on the sides of the cone $|F_t|=\mu F_n$, and  
correspond to times when the contact is sliding with
$F_t=\mu F_n$ or $F_t=-\mu F_n$ respectively.  The trajectory is not
quite a trapezoid, because after one cycle, the point does not return
to $A$, but arrives at $A'$, a bit closer to the origin than, but 
very close to, $A$.  After the following cycle, the system has again shifted 
towards the origin by the same amount.  This shift has its origin in
the tiny displacement that occurs with each cycle - the sliding contact
is gradually opening.

\begin{figure}[tb]
\centering
\includegraphics[height=0.5\textwidth,angle=-90]{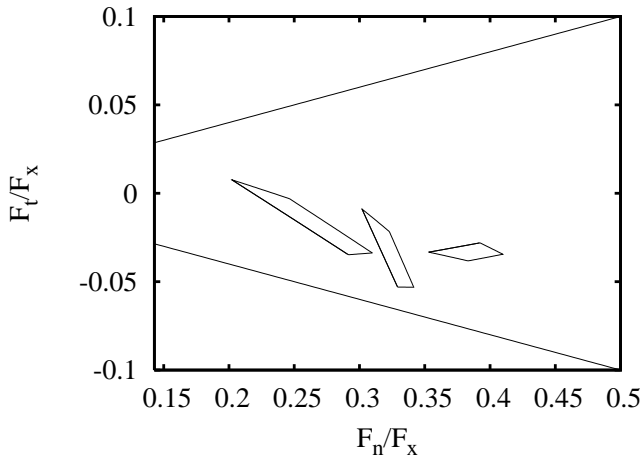}
\caption{Trajectories of selected non-sliding contacts in the
system shown in Fig.~\ref{stress-strain}.}
\label{NScontacts}
\end{figure}

The non-sliding contacts in the packing also trace out trapezoids,
but their edges do not intersect the cone $|F_t|=\mu F_n$.  Some
examples are shown in Fig.~\ref{NScontacts}.  
The corners of these trapezoids correspond
to the times when the sliding contact begins or stops sliding.
When all contacts remain non-sliding, the trajectories are no longer
trapezoids, but straight lines: under loading, each contact force moves
on a straight line, and under unloading, it simply retraces its path.
The reason for this is given in the appendix.

\subsection{Ratcheting without sliding contacts}
\label{sec:ellipse}

\begin{figure}[tb]
		\centering
		\includegraphics[width=0.5\textwidth]{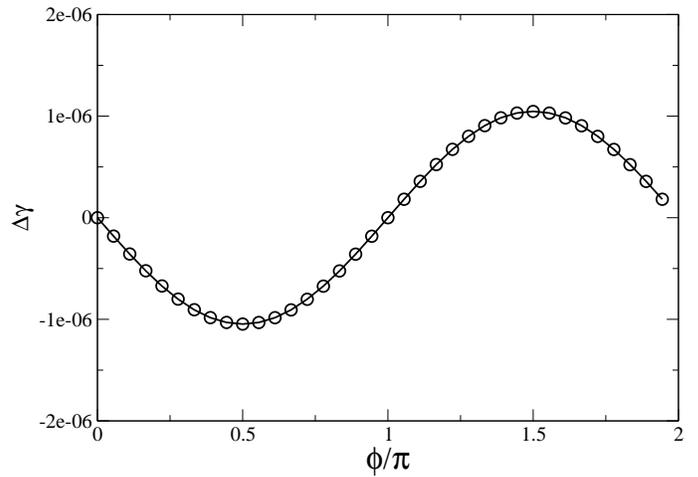}
		\caption{Strain per cycle under elliptic cyclic loading [see 
		Eqs.~(\ref{2Dforcing}) and (\ref{2Dqs})].  The circles are the observed points,
		and the line is a fit of the form $\Delta\gamma=A\sin\phi$.
		The simulations were done with 16 particles.  Sliding
		contacts were surpressed by setting $\mu=\infty$.}
		\label{fig:ellipse}
\end{figure}

\begin{figure}[tbp]
\centering
\includegraphics[width=0.5\textwidth]{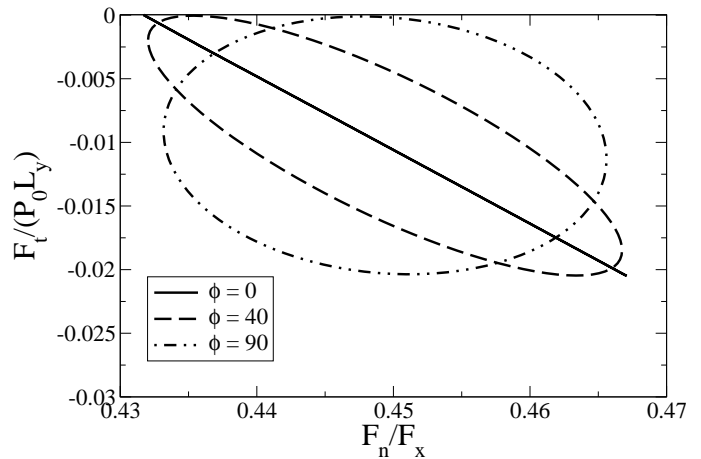}
\caption{Trajectory of a contact in the $(F_n,F_t)$ plane under
elliptic cyclic loading, for different values of the phase shift
$\phi$ (given in degrees).}
\label{fig:contact_ell}
\end{figure}

When the force on \textsl{both} walls is varied cyclically:
\begin{equation}
F_x = L_y \left[P_0 + q_x(t)\right], \quad F_y = L_x \left[P_0 + q_y(t)\right],
\label{2Dforcing}
\end{equation}
ratcheting can occur without sliding contacts. 
We carried out simulations of a small ($16$ particles) system with
Eq.~(\ref{2Dforcing}), with
\begin{equation}
		q_x(t) = \Delta\sigma(1-\cos\omega t+\phi), \quad
		q_y(t) = \Delta\sigma(1-\cos\omega t).
		\label{2Dqs}
\end{equation}
Note the presence of a phase shift $\phi$ between $F_x$ and $F_y$.
We call this form of loading \textsl{elliptic} cyclic loading, because
ellipses are traced out in the $(F_x,F_y)$ plane.
During these simulations, sliding
was suppressed by setting $\mu=\infty$.  
The results are shown in Fig.~\ref{fig:ellipse}.  The strain $\Delta\gamma$
per cycle is proportional to $\sin\phi$.  If one traces out the path
of $q_x(t),q_y(t)$ in the $q_x,q_y$ plane, then one obtains an ellipse
whose area is proportional to $\sin\phi$.  Tracing out any contact in
the $F_n,F_t$ plane also yields an ellipse proportional to $\sin\phi$.
Some examples are shown in Fig.~\ref{fig:contact_ell}.
This suggests that ratcheting is related to the area enclosed
by trajectories in the $(F_n,F_t)$ plane.

\subsection{Sign of the strain}
\label{sec:strainsign}

Ratcheting with small numbers of particles has another 
distinguishing property: the strain accumulation can be either
positive or negative.  Note that in Eq.~(\ref{imposed}), the average
imposed force does \textsl{not} correspond to an isotropic pressure,
because $q(t)\ge0$.  The pressure exerted by the walls on
the top and bottom of the sample are larger than at the side walls.
Thus one expects the sample to be gradually flattened, with the
top and bottom walls moving toward each other, while the side walls
are pushed apart.  This corresponds to $\gamma<0$ in Eq.~(\ref{defstrain}).
A series of $100$ different ratcheting simulations with $16$ particles
were performed, differing from each other only in the initial condition.
Of these $100$ simulations, $71$ exhibited unambiguously ratcheting.
Of these $71$ cases ratcheting, $51$ had $\gamma<0$ as expected, but
the remaining $20$ had $\gamma>0$.

On the other hand, when the sample size is larger, one has $\gamma<0$
whenever there is ratcheting.  A second series of $25$ simulations,
this time with $400$ particles, yields $18$ unambiguously ratcheting
simulations, all with $\gamma<0$.  The range of $\gamma$ observed
is also much smaller than for $16$ particles.  These results
suggest that the strain accumulated by a large sample is some kind
of average over strain accumulated by the small regions composing it.
In these small regions, there can be either negative or positive strain,
but after averaging, these fluctuations are smoothed out, so a large sample
has a quite predictable behavior.

Previous studies of granular ratcheting always considered large numbers
of particles, so this was never noticed.  

\section{Origin of Ratcheting}
\label{origin}

\subsection{Particle interaction model}
\label{sec:interaction}

\begin{figure}[tbp]
\centering
\includegraphics[width=0.33\textwidth]{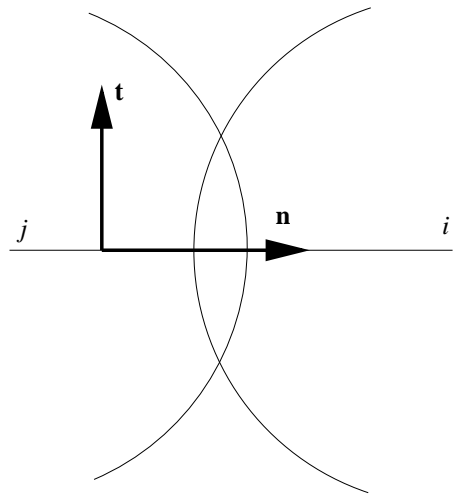}
\caption{Definitions for particle interaction laws.  The unit vector
$\bn$ defined in Eq.~(\ref{eq:defnormal}) points from particle $j$ toward
particle $i$, and $\bt=\mathbf{z}\times\bn$, where the $z$ axis points
out of the page.}
\label{fig:two}
\end{figure}

We now turn our attention from the description of granular ratcheting
to its cause.  We will consider
a non-sliding contact between two particles subjected to cyclic
external forces.  To facilitate the analysis, we assume that
$F_n$ is linear in the overlap distance.  One imagines that when two
grains first touch, two springs are created, one in the tangential and
the other in the normal direction.  Both springs obey Hooke's law 
so that the normal and tangential contact
forces are proportional to the spring elongations $D_n$, $D_t$:
\begin{equation}
F_n = -K_n D_n, \quad F_t = -K_t D_t,
\label{Feqn}
\end{equation}
where $K_n$ and $K_t$ are the spring constants.  Here, $F_n>0$ is interpreted
as pushing the particles apart, and $D_n<0$ occurs when the particles
overlap.  Eq.~(\ref{Feqn}) holds only for touching $D_n<0$ particles,
so $F_n>0$ in accord with Eq.~(\ref{contactconditions}).  On the other
hand, $D_t$ can have either sign, corresponding to the two
opposite tangential directions (up and down in Fig.~\ref{fig:two}).

The springs are stretched by the relative motion of the particles,
as long this does not violate any of the conditions in 
Eq.~(\ref{contactconditions}).  When the contact is non-sliding,
one has
\begin{equation}
\frac{dD_n}{dt} = V_n, \quad
\frac{dD_t}{dt} = V_t,
\label{DeqnNS}
\end{equation} 
where $V_n$ and $V_t$ are just the relative velocities at the point
of contact:
\begin{eqnarray}
V_n &=& (\mathbf{v}_i - \mathbf{v}_j)\cdot \bn , \\
V_t &=& (\mathbf{v}_i - \mathbf{v}_j)\cdot \bt  
  - r_i \omega_i - r_j \omega_j,
\label{eq:Vt}
\end{eqnarray}
where $\mathbf{v}_i$, $\omega_i$, and $r_i$ are the velocity, angular
velocity and radius of particle $i$, and $i$ and $j$ label the
touching particles.  The unit vector $\bn$
\begin{equation}
\mathbf{n} = \frac{\mathbf{x}_i - \mathbf{x}_j}{|\mathbf{x}_i - \mathbf{x}_j|}
\label{eq:defnormal}
\end{equation}
points from particle $j$
toward particle $i$, and $\bt$ is a tangent vector.  If the two-dimensional
space is assumed to be embedded in a three dimensional one, $\bt$ can be
defined as $\bt = \mathbf{z}\times\bn$, as shown in Fig.~\ref{fig:two}.
The forces $F_n$ and $F_t$ are then directed along $\bn$ and $\bt$
respectively.  Note that the signs in Eq.~(\ref{eq:Vt}) depend on the
choice of $\bn$, $\bt$, and the meaning of positive and negative 
$D_t$.  In Fig.~\ref{fig:two}, $D_t>0$ means points attached to particle
$i$ move upward relative to points attached to particle $j$.

If a contact opens, then
$F_n=F_t=0$ in accord with the first condition of 
Eq.~(\ref{contactconditions}).  If two separated particles come together
again, there is no memory of the previous contact.  

The second condition in Eq.~(\ref{contactconditions}) is enforced by
setting
\begin{equation}
D_t = \pm \mu \frac{K_n}{K_t} D_n
\label{eq:Dtslide}
\end{equation}
whenever using Eq.~(\ref{DeqnNS}) would lead to a violation of 
Eq.~(\ref{contactconditions}).

Sliding contacts are accounted for by modifying Eq.~(\ref{DeqnNS}),
but we do not need to consider this in detail, since sliding is
not needed for ratcheting to occur, as shown in Sec.~\ref{sec:ellipse}.

Note that no damping has been included in Eq.~(\ref{Feqn}).  This is
because ratcheting is a quasi-static phenomenon.  As the frequency of the
cyclic loading becomes very long, the deformation per cycle
approaches a constant.
In the limit of an infinitely long cycle, the particle
velocities vanish.  Any damping will also vanish, since it is proportional
to the velocities.  Since ratcheting exists in the limit of infinitely long
cycles, one does not need to consider damping in order to understand
granular ratcheting.  Damping is always included in simulations to
model the loss of energy when grains collide or slide against one another.

The model that has been described above has been in use for almost
thirty years \cite{Cundall79}.  It has been used in many different studies,
and considered to be well understood.  Nevertheless, we show that
this model contains an approximation that generates granular ratcheting.

\subsection{Path dependent potential energy}
\label{inequality}

Granular ratcheting occurs because the model described in
Sec.~(\ref{sec:interaction}) yields a path-dependent potential
energy.  Here, we are referring to the potential energy stored in
a contact when two particles overlap.  It models the elastic energy
stored when two grains are pushed together.  If the force pushing two
particles together is suddenly released, this potential energy is
converted into kinetic energy, and the particles will separate.  When
they separate, the highest possible kinetic energy they can attain is
\begin{equation}
E = \frac12 (K_nD_n^2 + K_t D_t^2).
\label{eq:PE}
\end{equation}
Thus Eq.~(\ref{eq:PE}) gives the potential energy stored in the
contact.

\begin{figure}[tb]
\centering
\includegraphics[width=0.5\textwidth]{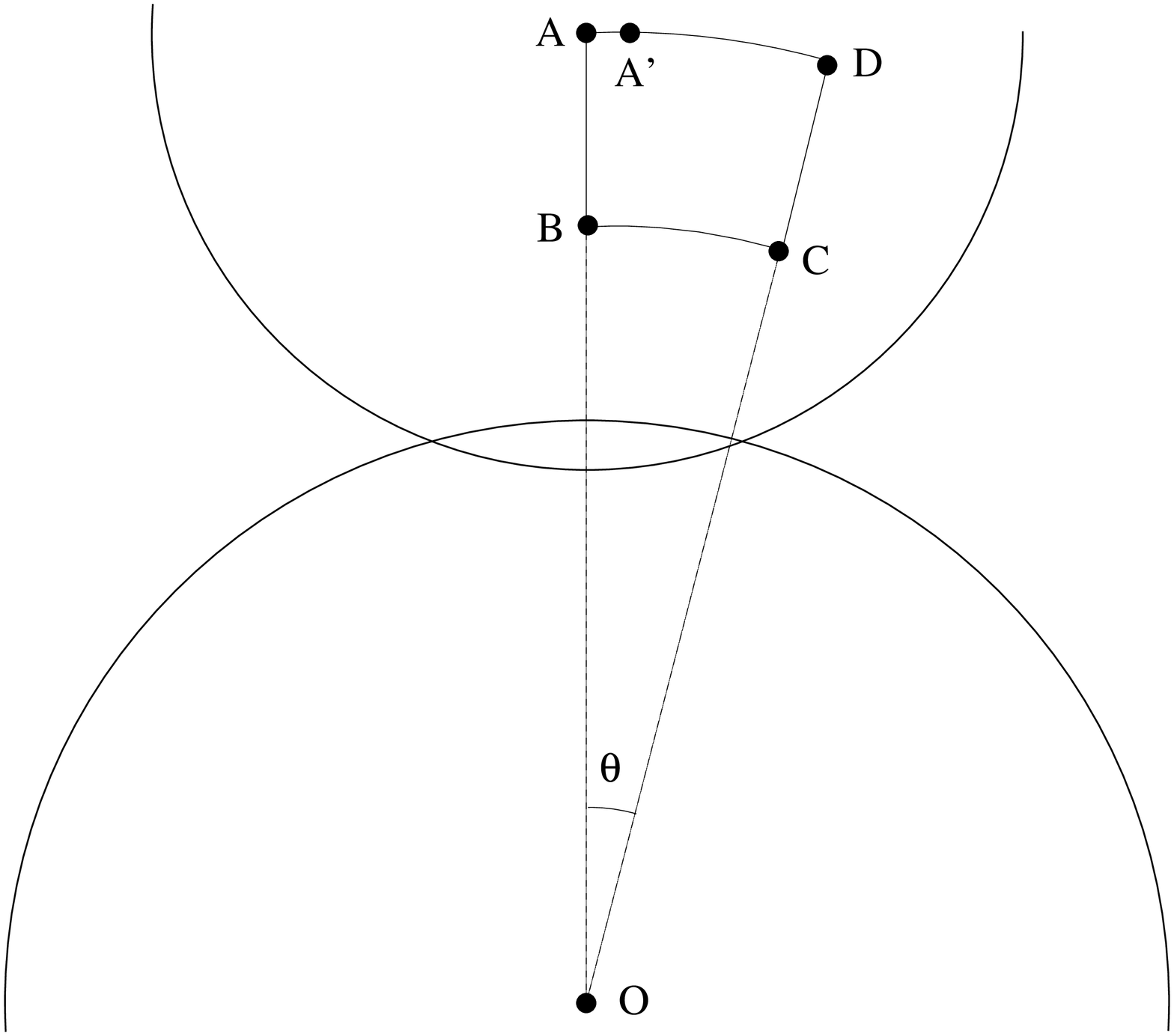}
\caption{The origin of granular ratcheting.  This figure shows two touching
disks.  The lower disk is fixed, and the upper disk moves without
rotating, its center tracing out the closed path $A\to B \to C \to D \to A$.
The contact
forces return to their initial state only if the upper particle stops at
$A'$ instead of proceeding to $A$.}
\label{originfig}
\end{figure}

We now show that this energy can be changed if the particles execute
a closed path relative to one another.  Consider the path shown in
Fig.~\ref{originfig}.  This figure shows a single contact
between two particles.  Let the lower particle be fixed,
and let the contact between the two grains be always non-sliding.
The point $A$ marks the center of the upper
particle, which is then moved so that it
traces out the path: $A\to B \to C \to D \to A$.
Neither particle rotates.  Even though the path is closed,
the length $D_t$ of the tangential spring is changed.

\begin{figure}[tb]
\centering
\includegraphics[width=0.5\textwidth]{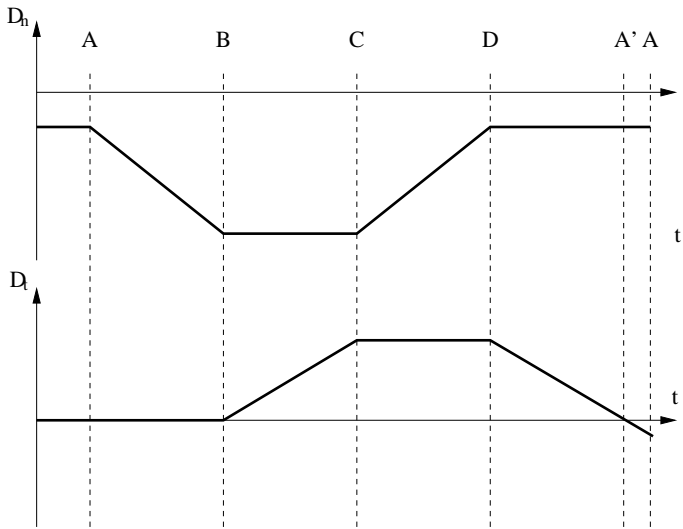}
\caption{Spring lengths $D_n$, $D_t$
for the cycle sketched in Fig.~\ref{originfig}.
Initially, $D_t = 0$ and $D_n<0$.  
As the upper disk moves from $A$ to $B$,
$D_n$ decreases.
Then $D_t$ increases
as the upper disk moves from $B$ to $C$. 
When it arrives at $D$,
$F_n$ has returned to its original value.
However, upon closing the cycle (returning to $A$),
$F_t$ \textsl{does not} return to its initial value.}
\label{forces}
\end{figure}

The changes in $D_n$ and $D_t$ during this cycle are sketched in
Fig.~\ref{forces}.
The segments $AB$ and $CD$ change only the normal spring length $D_n$, whereas
the arcs $BC$ and $DA$ change only the tangential spring length $D_t$.
Segments $AB$ and $CD$ are of equal length, so at the end of the cycle, $D_n$
has returned to its initial value.  However, arc $BC$ is shorter than arc $DA$
because it lies closer to the center of the lower particle.  Therefore, $D_t$
does \textsl{not} return to its original value, because Eq.~(\ref{eq:Vt})
implies that the change in the tangential spring length depends only on the
distance moved, irrespective of the distance between the touching particles.
Thus a cycle that returns the particles to their initial positions can modify
the potential energy.  The potential energy of a contact does not depend
only on the coordinates of the grains, but also on the past relative movements.

To see why this leads to granular ratcheting, note that $D_t$ determines
not only the potential energy, but also the tangential force.
Thus, when the particle executes the cycle shown in Fig.~\ref{originfig},
and returns to $A$, the contact force has also been modified.

Now let us consider a packing of particles, subjected to quasi-static cyclic
loading.  At the beginning of the loading cycle, the packing is in static
equilibrium, so that the net force on each particle vanishes.  
As the external load is varied, the contact forces and the particle
positions must also change.  After one loading cycle, the external load
has returned to its initial value.  If all the particles return to their
initial positions and all the contact forces to their initial values,
then there is no deformation of the sample, and thus no ratcheting.
On the other hand, if the contact forces have not returned to their
initial values, the packing will no longer be in force equilibrium,
and some deformation must occur.



\subsection{The role of sliding contacts}
\label{slidesection}

\begin{figure}[tb]
\centering
\includegraphics[width=0.5\textwidth]{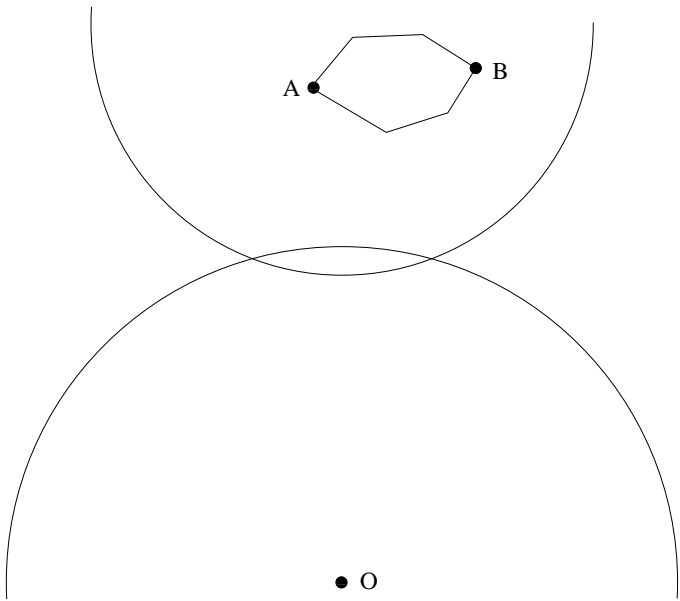}
\caption{A cycle where the lower particle is fixed and the upper
particle traces out a convex polygon.  If the trajectory is given
in polar coordinates with the origin being the center $O$ of the
lower particle, points $A$ and $B$ are where the angular coordinate
takes on its maximum and minimum values.}
\label{convex}
\end{figure}

The explanation of ratcheting presented here makes no reference to
sliding contacts.  Yet earlier studies identified sliding contacts as
a requirement for ratcheting.
To understand the
role of sliding contacts, it is necessary to consider a more
general motion, such as the one shown
in Fig.~\ref{convex}, where the upper particle
traces out a convex polygon.  The trajectories of the non-sliding contacts
shown in Fig.~\ref{NScontacts} are possible examples.
Again let us use polar coordinates, with
the origin placed at $O$.  The path of the upper particle will now
be given by $r(t)$ and $\theta(t)$, where $t$ is time.
We identify two points, labeled $A$ and $B$
in the figure, where $\theta(t)$ attains its maximum and minimum values.
At any time, the tangential velocity is given by
\begin{equation}
V_t = r \frac{d\theta}{dt},
\label{Vtpolar}
\end{equation}
and so the total change of the tangential spring, as the particle
moves from $A$ to $B$ is
\begin{equation}
\Delta D_t(A \to B) = \int_{t_A}^{t_B} r(t)  \frac{d\theta}{dt}\,dt = 
   \int_{\theta_A}^{\theta_B} r_{AB}(\theta)\,d\theta,
\label{AB}
\end{equation}
where $r_{AB}(\theta)$ gives the trajectory that the particle follows
from $A$ to $B$.  The change in $D_t$ on the return trip is
\begin{equation}
\Delta D_t(B\to A) = \int_{\theta_B}^{\theta_A} r_{BA}(\theta)\,d\theta,
\label{BA}
\end{equation}
where $r_{BA}(\theta)$ is the path followed from $B$ back to $A$.
The total change in length of the tangential spring is obtained
by adding Eqs.~(\ref{AB}) and (\ref{BA}) together:
\begin{equation}
\Delta D_t = \int_{\theta_A}^{\theta_B} \left[r_{AB}(\theta)-r_{BA}(\theta)\right] d\theta.
\label{creep}
\end{equation}
If the particles are very stiff, then the deformations are small:
$|\theta_A-\theta_B|\ll 1$ and
$|r_{AB}-r_{BA}| \ll r_i+r_j$.  Then Eq.~(\ref{creep}) can be written
\begin{equation}
\Delta D_t = \frac{a}{r_i +r_j},
\label{eq:area}
\end{equation}
where $a$ is the area enclosed by the trajectory of upper particle.

Now the role of the sliding contacts becomes clear.  If there are
no sliding contacts, then the trajectories are straight lines, and
$r_{AB}(\theta)=r_{BA}(\theta)$ for all
$\theta_B \le \theta \le \theta_A$, and $a=0$.  Thus there is no change
in $D_t$ if the particles return to their original positions, and
ratcheting does not occur.  On the other hand, the presence of
sliding contacts guarantees that $a\ne0$, so the particles cannot
return to their original positions, and ratcheting occurs.  The reason
why sliding contacts are required to obtain trajectories that enclose a
non-zero area is explained in Sec.~\ref{sec:stiffness}.

\section{Angular Molecular Dynamics}

\subsection{Algorithm}
\label{sec:AMD}

\subsubsection{Definition of the tangential spring}
\label{sec:newTspring}

To confirm our explanation of granular ratcheting, we show how it
can be eliminated by using a new method of calculating the tangential 
forces where the potential energy is path-independent.  To do
so, we retain Eq.~(\ref{eq:PE}), but define $D_n$ and $D_t$ in such
a way that they depend only on the coordinates of the particles.
For the spring in the normal direction, this is
straightforward.  If the particle positions are given, the
overlapping distance can be used as the normal spring length:
\begin{equation}
D_n = r_i+r_j - \left|\bx_i - \bx_j \right|.
\label{eq:DNposition}
\end{equation}
where $\bx_i$ and $\bx_j$ are the positions of the touching
particles and $r_i$ and $r_j$ their radii.

\begin{figure}[tbp]
\centering
\includegraphics[width=0.33\textwidth]{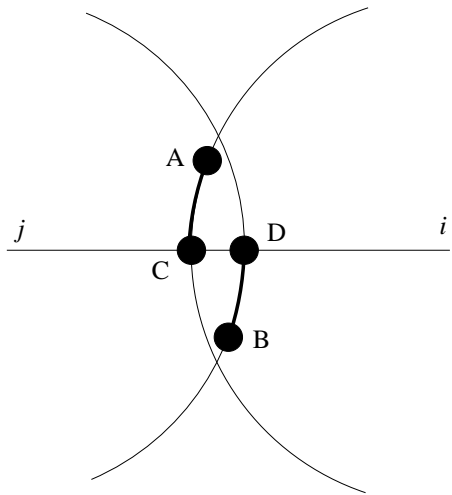}
\caption{The definition of the tangential spring.  Its length is equal
to the arc length $AC$ plus the arc length $DB$.  Points $A$ and $B$
are defined when the two particles first touch.  They are carried by
the rigid body motion of the particles.  
Points $C$ and $D$ are defined by the intersection
of the the line connecting the centers (horizontal line) 
with the particle surfaces.}
\label{fig:AMD1}
\end{figure}

For the tangential spring, the point of first contact must be stored.
Let us imagine that when two particles first touch, a spot is painted on
each particle, marking the point where they touch.  Let these points
be called $A$ and $B$.  The points of first contact
are fixed to the particle surfaces, and thus
carried with the subsequent solid-body motion of the
particles.  To determine the tangential spring length at a later time,
one first determines the current points of contact $C$ and $D$.
These points are defined by the intersection of the particle surfaces
with the line connecting the centers.  The tangential spring length is
the length of the arc $AC$, plus the length of the arc $DB$,
as shown in Fig.~\ref{fig:AMD1}. 

\begin{figure}[tbp]
\centering
\includegraphics[width=0.33\textwidth]{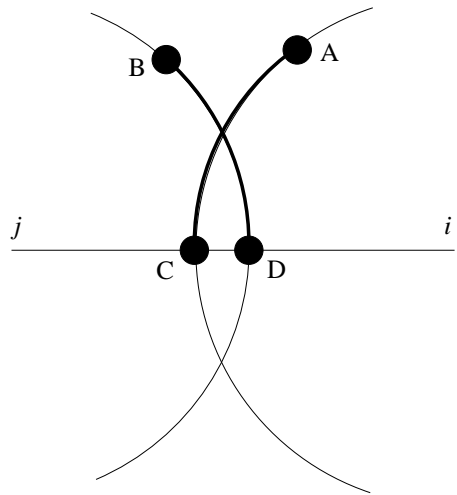}
\caption{Definition of rolling distance. 
It is the average
of the distance $AC$ and the distance $BD$.
The points are defined as in Fig.~\ref{fig:AMD1}.}
\label{fig:AMD2}
\end{figure}

One useful side effect of calculating the tangential springs in this way is
that one can easily obtain the distance the particles roll relative to one
another.  If two particles touch, and then roll without sliding,
the points $A$ through $D$ will be as sketched in Fig.~\ref{fig:AMD2}.
The distance rolled is the length of the arc $AC$ or $BD$
(see Fig.~\ref{fig:AMD2}).  Note that this measure of the rolling is objective,
because it is based on points fixed on the particles themselves, and is thus
independent of any solid-body motion imposed on the two particles.
 
Each point of a circle can be assigned an angle, obtained
by measuring the angle between the $x$-axis and 
a line determined by the point in question
and the center of the circle. 
In this way one can assign the angles $\theta_A$, $\theta_B$, $\theta_C$,
and $\theta_D$ to the points $A$, $B$, $C$, $D$, and arc lengths can now
be calculated by subtracting angles.  Thus the arc length
$AC$ is $r_i(\theta_A-\theta_C)$.  Note that the arc length has
a sign, which is necessary for distinguishing between rolling and
sliding.  Note also that $DB$ is $r_j(\theta_B-\theta_D)$ since
angles are measured with respect to particle $j$ and not $i$.

Now we can write the tangential spring length as
\begin{equation}
D_t = r_j(\theta_C-\theta_A) + r_i(\theta_D-\theta_B),
\label{eq:AMDtangent}
\end{equation}
where we adopt the convention that $D_t$ increasing means that point
$A$ in Fig.~\ref{fig:AMD1} moves upward (i.e., $\theta_A$ decreases),
and $B$ moves downward (i.e., $\theta_B$ decreases).
The rolling distance is
\begin{equation}
D_\mathrm{roll} = \frac{r_j(\theta_C-\theta_A) - r_i(\theta_D-\theta_B)}{2}
\label{eq:AMDroll}
\end{equation}

\subsubsection{Direct implementation}
\label{sec:AMDpure}

The most obvious way to implement this algorithm is to calculate
the angles $\theta_A$ through $\theta_D$, and then use 
Eq.~(\ref{eq:AMDtangent}) to calculate the spring length.

The angles $\theta_A$ and $\theta_B$ can be found by integrating
the equations
$\dot\theta_A = \omega_j$, $\dot\theta_B = \omega_i$.
But 
it may be more economical to assign an angular coordinate $\theta_i(t)$
to each particle $i$.  When the particle positions are updated,
$\theta_i$ can be updated as well, using $\dot\theta_i = \omega_i$.
Then at the time $t_*$ of first contact, one stores $\theta_i(t_*)$,
and $\Delta\theta_i = \theta_i(t_*) - \theta_n$, where $\theta_n$
is the angle of the point of first contact at time $t_*$.  Then
at any later time $t$:
\begin{equation}
\theta_A = \theta_i(t) - \theta_i(t_*) + \Delta\theta_i.
\label{eq:thetaA}
\end{equation}

The angles $\theta_C$ and $\theta_D$ are calculated at each time step
from the positions of the particles.  Writing
$n_x$ and $n_y$ for the two components of $\bn$, 
\begin{equation}
\theta_C = \left\{ \begin{array}{cc}
\tan^{-1}n_x/n_y, & n_x > 0,\\
\pi-\tan^{-1}n_x/n_y & n_x < 0.
\end{array}\right.
\label{eq:thetaC}
\end{equation}
Then $\theta_C$ is moved into the correct interval by adding or subtracting
$2\pi$.  Then one uses $\theta_D = \theta_C \pm \pi$.
In this way, both the rotation
and translation of the particles is taken into account.

If a contact slides, one moves the points $A$ and $B$ along the particle
surfaces so that Eq.~(\ref{eq:Dtslide}) is satisfied.  In a similar way,
one could move these points to set $D_\mathrm{roll}=0$ while
leaving $D_t$ unchanged.

\subsubsection{Implementation through integration}
\label{sec:CMDcorr}

An alternative way of implementing this algorithm
is to modify the existing Cundall-Strack algorithm. 
A few minor modifications are necessary
to obtain an algorithm with a potential energy
given in Eq.~(\ref{eq:PE}) which is path-independent,
with $D_t$ defined as in Eq.~(\ref{eq:AMDtangent}).

To do this, first
express $\dot D_t$ and $\dot D_\mathrm{roll}$ in terms
of the motion of the particles.  To do this, we need
\begin{equation}
\dot{\bn} = \frac{\left(\mathbf{v}_i - \mathbf{v}_j\right)\cdot\bt}
   {|\mathbf{x}_i - \mathbf{x}_j|}\bt.
\label{eq:dndt}
\end{equation}
where the tangent vector is $\bt=\mathbf{z}\times\bn$
as in Fig.~\ref{fig:two}.
Then Eqs.~(\ref{eq:AMDtangent}), (\ref{eq:thetaA}), and (\ref{eq:thetaC}) give
\begin{equation}
\dot D_t = - r_i\omega_i - r_j\omega_j + \alpha
(\mathbf{v}_i-\mathbf{v}_j)\cdot \mathbf{t},
\label{eq:DerivDt}
\end{equation}
where we have defined
\begin{equation}
\alpha = \frac{r_i + r_j}{|\mathbf{x}_i - \mathbf{x}_j|}
\label{eq:defalpha}
\end{equation}
Note that the first of these is equivalent to Eq.~(\ref{DeqnNS}) and
(\ref{eq:Vt}) only when $\alpha=1$ or
$|\bx_i-\bx_j|=r_i+r_j$, i.e. when the particles are just touching.
The usual implementation of the Cundall and Strack model, therefore,
contains an approximation, namely $\alpha \approx 1$.  Normally
one chooses a stiffness high enough so that this approximation is
reasonable, but it nevertheless has an effect on the simulation
results.

In the same way, one can obtain a rolling velocity from
Eq.~(\ref{eq:AMDroll}):
\begin{equation}
\dot D_\mathrm{roll} = - r_i \omega_i + r_j \omega_j + 
\frac{r_i-r_j}{|\mathbf{x}_i - \mathbf{x}_j|}(\mathbf{v}_i-\mathbf{v}_j)
\cdot \mathbf{t},
\label{eq:DerivDroll}
\end{equation}

To obtain the equations of motion,
one cannot simply use Eq.~(\ref{Feqn}).  To guarantee conservation
of energy, one defines the Lagrangian \cite{Goldstein}
\begin{equation}
L = T - V
\label{eq:Lagrangian}
\end{equation}
where $T$ is the kinetic energy of a system, and $V$ is the potential
energy.  In our case, we consider the two touching particles whose
kinetic energy is
\begin{equation}
T = \frac12 m_i \left(\dot x_i^2 + \dot y_i^2\right) + 
    \frac12 m_j \left(\dot x_j^2 + \dot y_j^2\right)
 + \frac12 I_i \dot \theta_i^2 + \frac12 I_j \dot \theta_j^2.
\label{eq:KE}
\end{equation}
The potential energy $V$ is given by Eq.~(\ref{eq:PE}).  The equations of motion
are then given by
\begin{equation}
\frac{d}{dt}\left(\frac{\partial L}{\partial \dot q} \right )
	- \frac{\partial L}{\partial q} = 0,
\label{eq:eqmotion}
\end{equation}
where $q$ is one of the coordinates of the grains $x_i, y_i, \theta_i,
x_j, y_j, \theta_j$.  Applying this equation yields
\begin{eqnarray}
m_{i,j} \mathbf{\ddot x}_{i,j} &=& \pm K_n D_n \mathbf{\hat n}
	\pm \alpha K_t D_t \mathbf{\hat t},\cr
I_{i,j} \ddot \theta_{i,j} &=& K_t D_t r_{i,j}.  
\label{eq:eqmotion2}
\end{eqnarray}
Note that these differ from Eq.~(\ref{Feqn}) by the presence of the factor
$\alpha$ in the tangential force.  This same factor appears in
Eq.~(\ref{eq:DerivDt}).
This means that this new method can be implemented simply by inserting
this factor in the appropriate places in the program.

\subsection{Results}

We have compared the traditional Cundall-Strack algorithm used in
Secs.~\ref{sec:ellipse} and \ref{sec:strainsign} with the
the two different implementations of the
angle-based algorithm discussed in Sec.~\ref{sec:AMD}.

\subsubsection{Simulation Parameters}

In all cases, the initial condition was generated by placing
grains on a lattice in a square domain.  The radii are uniformly
distributed within the interval $r_\mathrm{max}[0.7,1]$, where
$r_\mathrm{max}$ is chosen so that the desired number of particles
will fit in the domain.  

Two walls of the domain are fixed, and the other two are movable.
A force proportional to wall length is applied to the movable walls, and
they compress the particles at uniform stress into a packing.
During this time of compression, the particles are smooth
(friction ratio $\mu=0$).  Once this compression is complete,
one sets $\mu=0.2$, and imposes cyclic loading as described above.

The system of units for the simulation is given by the initial length $L$
of the system, the (two-dimensional) pressure $p$ applied during compression,
and the density $\rho$ of the particles.   In these units, the stiffness
of the particles is $K_n = K_t = 100p$.  The unit of time is
$\tau = \sqrt{m/p}$.  One cycle lasts $10\tau$.  At least $100$ cycles were
performed in all simulations.

The position of the movable walls are
recorded at the time of minimum force during each cycle.  By comparing
these values from one cycle to the next, an accumulation of strain
can be detected.  To determine whether a sample ratchets, the following
procedure was applied.  First, the first $29$ cycles were neglected
to eliminate transients.  Then $L_{x0}$ and $L_{y0}$ were defined
by the positions of the walls at the beginning of the thirtieth cycle.
Next, the strain $\gamma$, defined in Eq.~(\ref{defstrain}) was calculated
for each subsequent cycle.  Finally, we checked whether $\gamma$
increases linearly with cycle number $N$.  This was done by fitting a line
to the observed $(\gamma,N)$, and calculating
the root mean square deviation of the observed points from the
fit.  If this number was smaller than the slope,
the simulation was judged to exhibit ratcheting.  Otherwise, it was
considered non-ratcheting.

\subsubsection{Ratcheting in small systems}

We subjected $100$ different small packings ($16$ particles) to cyclic
loading as described above.  With the unmodified Cundall-Strack algorithm,
$71$ simulations exhibited ratcheting.  The deformation per cycle
$\Delta\gamma$ varied over a wide range: $10^{-12}<|\Delta\gamma|<10^{-6}$,
with a geometric mean of $8\times10^{-9}$.
Both positive and negative values were observed: 
$-10^{-6}<\Delta\gamma<4\times10^{-7}$.

\begin{figure}[tbp]
\centering
\includegraphics[width=0.5\textwidth]{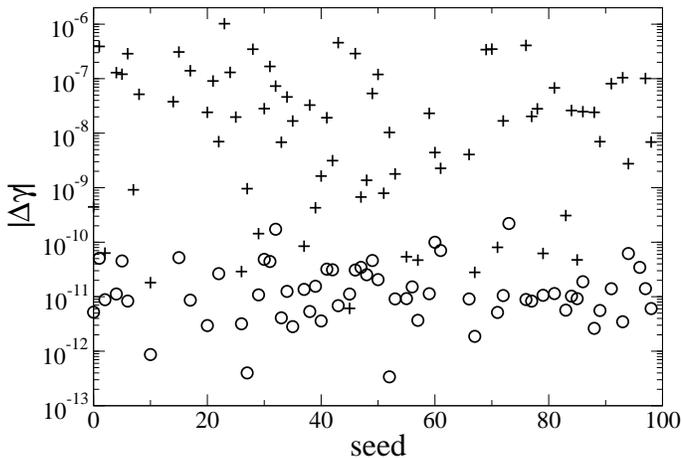}
\caption{Deformation per cycle for all ratcheting simulations.
$100$ different configurations were generated using different random
number seeds.}
\label{fig:CMDcompare}
\end{figure}

When the Cundall-Strack algorithm is modified as described in
Sec.~\ref{sec:CMDcorr},
$62$ still simulations exhibit ratcheting, but
at a much lower amplitude.  One observes
$10^{-13}<|\Delta\gamma|<2.5\times10^{-10}$ with a geometric mean
of $1.1\times10^{-11}$.  These results are summarized in
Fig.~\ref{fig:CMDcompare}.  One sees that the use of
the corrected equations leads to a $10^4$-fold reduction of granular
ratcheting.

\begin{figure}[tbp]
\centering
\includegraphics[width=0.5\textwidth]{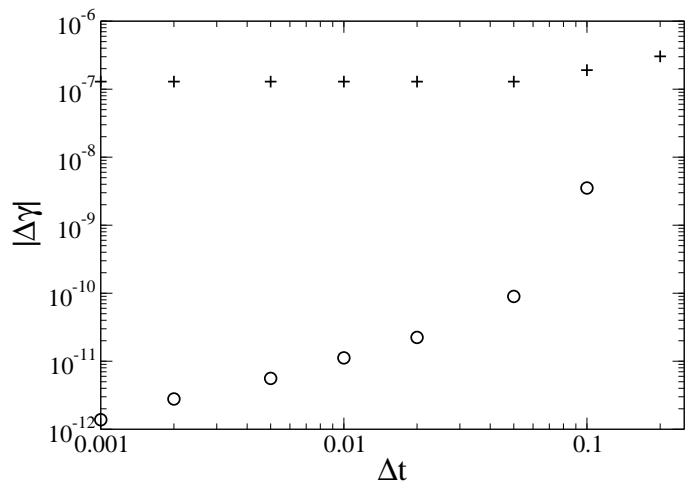}
\caption{Dependence of ratcheting on time step.  Crosses: original
algorithm.  Circles: corrected algorithm.  The time step is given
in multiples of $\sqrt{K/m_\mathrm{min}}$, where $m_\mathrm{min}$
is the smallest particle in the simulation.}
\label{fig:CMDdt}
\end{figure}

The remaining granular ratcheting is due to integration errors.
This can be shown by taking a single initial condition, and changing
the time step.  Typical results are shown in Fig.~\ref{fig:CMDdt}.
With the original Cundall-Strack algorithm, ratcheting is independent
of the time step.  When it is modified, then the ratcheting deformation
is proportional to the time step. 

Finally, when the method described in Sec.~\ref{sec:AMD} is implemented
by direct calculation of angles, no simulations ratchet.  $28$ of
the simulations exhibit a constant strain with $|\gamma|<10^{-14}$
for every cycle.  Note that such small deformations are not even
visible on Figs.~\ref{fig:CMDcompare} and \ref{fig:CMDdt}.
The others exhibit a variety of other behaviors that will be discussed
in the next section.

\subsubsection{New behaviors in small systems}

\begin{figure}[tbp]
\centering
\includegraphics[width=0.5\textwidth]{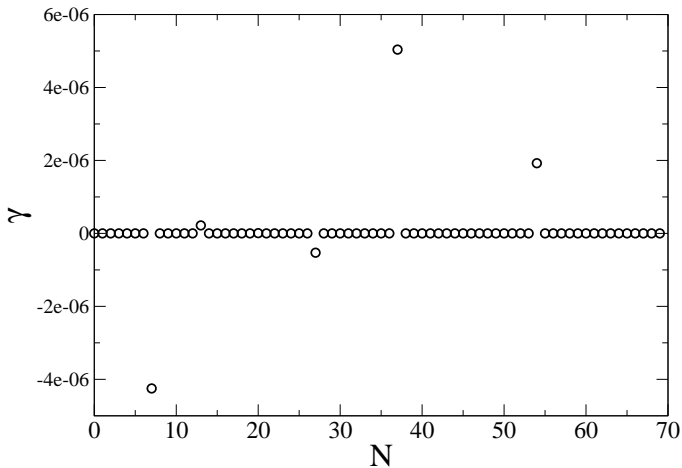}
\caption{Strain as a function of cycle number for a simulation
with outliers.}
\label{fig:outliers}
\end{figure}

Once ratcheting has been removed or reduced, new behaviors come to
the foreground.  The most common of these are outliers.  The strain
is independent of cycle number, except for occasional cycles.  An
example is shown in Fig.~\ref{fig:outliers}.

\begin{figure}[tpb]
\begin{center}
\includegraphics[width=0.5\textwidth]{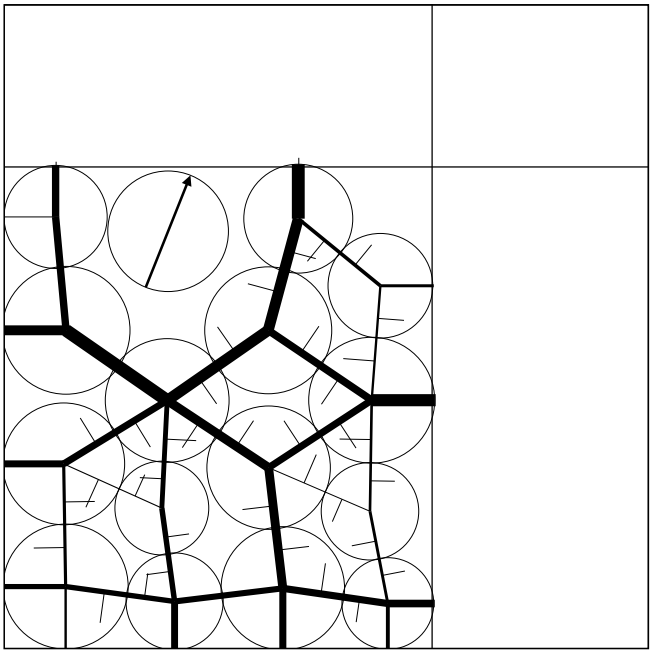}
\end{center}
\caption{A configuration that causes outliers.  The arrow shows the
motion of the rattler.  The normal forces are
shown by the lines connecting particle centers.  The tangential
forces are shown by lines perpendicular to the normal forces.
Note that the rattler does not participate in the force network. }
\label{fig:outlier_config}
\end{figure}

A closer inspection of these simulations reveals that these outliers
are due to ``rattlers'': particles without contacts.  Since there is
no gravity, rattlers float inside cages in the packing.  Occasionally,
they collide with walls of their cage.  These collisions coincide
with the outlying points.  Once the collision is past, the
packing returns to its initial state, and the rattler floats
off toward another part of its cage.  The packing that produced
Fig.~\ref{fig:outliers} is shown in Fig.~\ref{fig:outlier_config}.
The rattler is found in the upper center of the packing.  It moves
within a cage formed by five particles and the upper wall.  The
outlying points in Fig.~\ref{fig:outliers} coincide with collisions
between the rattler and its cage.  However, not all collisions leave
a trace in Fig.~\ref{fig:outliers}; the amplitude registered in
Fig.~\ref{fig:outliers} probably strongly depends on the time within
the cycle where the collision takes place.

\begin{figure}[tbp]
\centering
\includegraphics[width=0.5\textwidth]{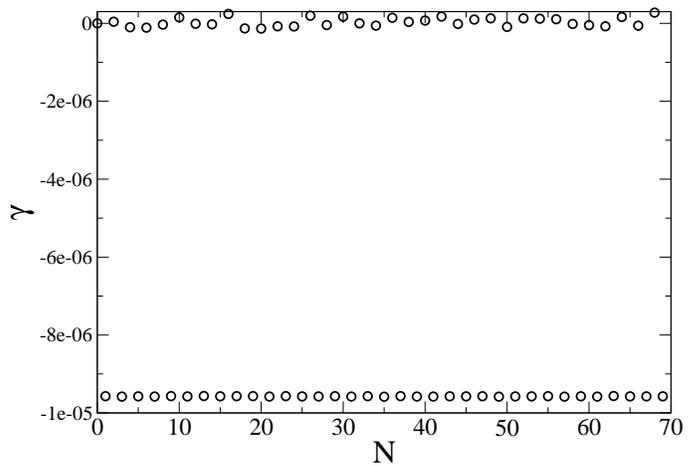}
\caption{Accumulated strain $\gamma$ versus cycle number for a simulation
performed with the angle based method of calculating tangential forces.
The strain is periodic.  This arises when the rattler is confined in a
very small cage.}
\label{fig:periodic}
\end{figure}

The frequency of collisions with the cage can vary widely.  Sometimes
only one point out of $70$ is perturbed. In other cases, every
point can be considered as an outlier.  Another thing that can
happen when the rattler's cage is small is that it can be driven
in the cage by the motion of the surrouding particles in a periodic
way.  This leads to a periodic dependence of $\gamma$ on $N$,
as shown in Fig.~\ref{fig:periodic}.

Rattler-induced outliers exist also in the original Cundall and Strack
method.  When on inspects the $29$ non-ratcheting simulations,
one finds that $25$ of them have
perturbations due to rattlers.  

Another effect that rattlers can cause is a sudden step in the
strain.  This occurs when 
the particles forming the cage have only weak contact forces.
The rattler can induce a sudden step in the strain by provoking a
small re-arrangement of these particles.

One way to reduce the effect of the rattlers 
is to apply a weak gravitational field
so that they can no longer float slowly 
from one side of their cage to another. 
When this is done, outliers still exist, 
but the perturbations they introduce are much smaller
-- $O(10^{-12})$ instead of $O(10^{-7})$.

\begin{figure}[tbp]
\centering
\includegraphics[width=0.5\textwidth]{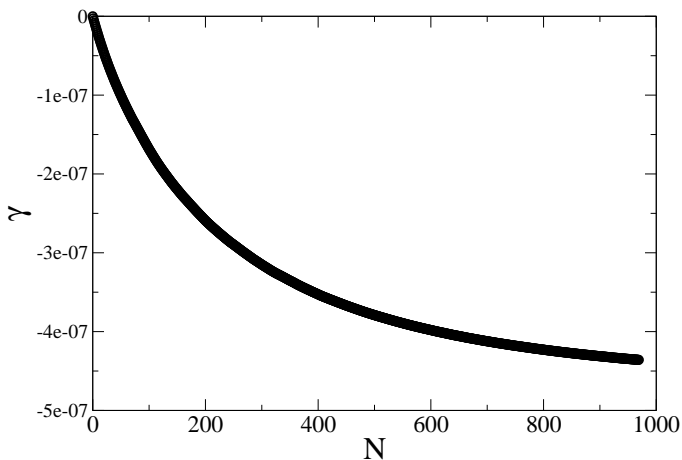}
\caption{Accumulated strain $\gamma$ versus cycle number for a simulation
performed with the angle based method of calculating tangential forces.
This sample exhibits shake down.}
\label{fig:shakedown}
\end{figure}

Another phenomenon is ``shakedown'', which has already been
investigated \cite{Ramon}.
In ``shakedown'', the accumulated strain per cycle
decreases each cycle.  In Fig.\ref{fig:shakedown}, 
we show an example.  The time
required to reach a level of negligible strain accumulation is
variable.  In Fig.~\ref{fig:shakedown}, strain is still accumulating, 
even after $1000$ cycles.  In other simulations, shakedown occurs after
very few simulations.

\subsubsection{Large systems}

To study these phenomena in larger systems, a series of $25$ simulations
with $400$ particles was carried out.  With the original Cundall--Strack
method, ratcheting is observed, with a much narrower range of strain
accumulation, as discussed in Sec.~\ref{sec:strainsign}.
When either of the modifications proposed in Sec.~\ref{sec:AMD} is used, 
outliers dominate the
stress - cycle graphs.  This is probably because as the packing becomes
large, the probability of having a rattler approaches unity.
When a weak gravitational field is applied, the phenomena of ``shakedown''
dominates.

\section{Conclusions}

We have uncovered the cause of granular ratcheting.  It is due
to a potential energy that depends not only on the particle positions,
but also on their past trajectories.  
As a granular assembly is subjected to cyclic
loading, it is impossible to return both the particle positions
and the contact forces
to their initial values, so a small deformation occurs with each cycle.
It follows that granular ratcheting can be eliminated by defining
a potential energy depending only on the current particle positions.
This was confirmed by extensive numerical simulations using
two different implementations of this idea.

This result suggests that contact modeling should focus on the
potential energy as the fundamental quantity, and then use
Eq.~(\ref{eq:eqmotion}) to obtain the forces.  
In contrast,
the most common approach taken in the literature
is to directly postulate forces based on physical grounds,
without considering the potential energy.

One possible criticism of this work is that it is only concerned with
disks, whereas ratcheting has been found in packings of polygons.
The motion of disks is much simper to analyze, because the rotation
of the disks is uncoupled from the translation.  In polygons, this
is no longer true.  But all that is required is that the potential
energy be path dependent.  One common force law, used in granular
ratcheting studies \cite{Fernando} is to assume that the force
between polygons is proportional to the overlap area.  This
force law has been shown to violate energy conservation \cite{Thorsten}.
Thus it seems likely that the explanation of granular ratcheting
presented here also applies to ratcheting of packings of polygons.

At a detailed level, the numerical mechanism cannot be the same as the
physical one.   Numerical granular ratcheting is a consequence
of the way the tangential spring is stretched.  In experiments, the
contacts between the particles are not governed by the stretching
of springs.  Indeed, if two touching particles can be considered as
making up a single elastic body, then force and position cycles will
coincide, as there is a potential energy.  

However, the results of this paper do show that granular ratcheting
in the experiments will occur if force and position cycles are not equal.  
In principle,
this could be checked by examining a single contact under cyclic loading.
Such an experiment would be difficult to do, since very small relative
displacements must be measured.  And it must also be mentioned that
the idea of two contacting particles acting as if they were
welded together at the contact surface is itself an idealization.
There may be
zones of slip at the contact (even when the contact as a whole does
not slide), and this may give rise to a complicated behavior when
the contact is subjected to cyclic loading.  Another possibility
is that fluid could coat the surfaces of the touching particles,
possibly lubricating them.  Or abrasion at the contact point could
generate very tiny particles trapped between the two touching surfaces.
These particles could act like fault gouge \cite{gouge} 
on a very small scale, facilitating a relative tangential motion.
All of these effects may lead to a history dependent potential energy,
and thus to granular ratcheting through the mechanism
discussed in this paper.


\bibliographystyle{prsty}
\acknowledgments

The authors acknowledge support from the Deutsche Forschungsgemeinschaft
through grant HE 2732/8-1 ``Mikromechanische Untersuchung des granularen
ratchetings'' and the German-Israeli Foundation (GIF).  
The authors thank Ciprian David for fruitful discussions.

\appendix
\section{Stiffness Matrix Theory}
\label{sec:stiffness}

The section presents a very brief review of stiffness matrix theory.
This theory applies to granular packings under quasi-static loading, 
and thus is applicable to granular ratcheting.  We explain below
how this theory explains certain key properties of packing under
cyclic loading, namely,
\begin{itemize}
\item why particle trajectories are straight lines when there are no
sliding contacts, and the forcing is given by Eq.~(\ref{imposed}),
\item why this is no longer true when there are sliding contacts, and
\item why the forcing given in Eq.~(\ref{2Dforcing}) generates
particle trajectories with a non-zero area.
\end{itemize}

\subsection{Introduction to stiffness matrix theory}

In stiffness matrix theory \cite{one}, the behavior of the packing is 
piece-wise linear.  Thus
time can be divided into intervals $[t_i,t_{i+1}]$ during which
the velocities of the particles are linearly related to the
change in forces:
\begin{equation}
\frac{d\mathbf{f}_\mathrm{ext}}{dt} = \mathbf{kv},
\label{PreviewStiffness}
\end{equation}
where $\mathbf{f}_\mathrm{ext}$ represents the external forces
($F_x$ and $F_y$ for the biaxial box), $\mathbf{v}$ contains
the velocities of the particles and walls, and $\mathbf{k}$ is called
the \textsl{stiffness matrix}.  It relates the velocities (or
displacement increments) of the particles to the change in the
force exerted on each particle by its neighbors.

The motion is only \textsl{piece-wise} linear
because the stiffness matrix $\mathbf{k}$ depends
on the status (sliding or not) of each contact.  
Whenever a contact status changes,
$\mathbf{k}$ also changes.
Therefore,
the times $\{t_i\}$ which define the intervals of linearity
are the times when one or more contacts change status.

Eq.~(\ref{PreviewStiffness}) holds when the forcing is quasi-static, and
the particles are quasi-rigid.  Quasi-static forcing means that the
time scale associated with the forcing is much longer than the time
the packing needs to react.  Particles can be said to be quasi-rigid if
their stiffness is much greater than the confining pressure.  Note that
these two assumptions are related: if the particles are very stiff, the
speed of sound is very high, and the packing can quickly react to
changes in the external load.

\subsection{Application to biaxial test}

If one considers a biaxial test, with the forcing given by 
Eq.~(\ref{imposed}), then only the entries
of $\mathbf{f}_{\mathrm{ext}}$ corresponding to the walls
are non-zero, because no external forces are applied to the particles.
Furthermore, in Eq.~(\ref{PreviewStiffness}), only those components
associated with varying forces survive differentiation by time.
Thus Eq.~(\ref{PreviewStiffness}) becomes
\begin{equation}
		\frac{dq}{dt}\mathbf{L}_x = \mathbf{kv},
		\label{BiaxialStiffness}
\end{equation}
where $\mathbf{L}_x$ is a vector, all of whose components are zero, except
the $y$ component of the force on the upper and lower walls.  All 
other components of $\mathbf{f}_\mathrm{ext}$ are either zero or constant.

One would like to invert $\mathbf{k}$
and bring it onto the left hand side of the equation. 
But $\mathbf{k}$ is singular
because there are certain collective motions
that do not change the spring lengths,
and hence do not change the forces. 
One example is the uniform motion of all particles.  
They do not move relative to one other,
and provoke no change in force.  
Let $\mathbb{B}$ be the set of all such motions. 
We can be sure that the left hand side of Eq.~(\ref{BiaxialStiffness})
is orthogonal to every member of $\mathbb{B}$,
for if it were not, the packing would be unstable \cite{one}.

Now define the matrix $\mathbf{\hat k}$ that will act like an inverse
of $\mathbf{k}$.  It is defined by
\begin{equation}
\mathbf{kx}=\mathbf{f} \mbox{ and } \mathbf{x}\perp\mathbb{B}
\; \Rightarrow \; \mathbf{\hat kf}=\mathbf{x}.
\end{equation}
This equation gives the result of applying $\mathbf{\hat k}$ for 
$3N-\dim \mathbb{B}$ linearly independent vectors.  To fully determine
$\mathbf{\hat k}$, we must say how it acts on the other $\dim \mathbb{B}$
dimensions of $\mathbb{R}^{3N}$.  Let $\mathbb{F}$ be the range of 
$\mathbf{k}$.  Then:
\begin{equation}
\mathbf{\hat kf}=0, \mbox{ for } \mathbf{f} \perp \mathbb{F}.
\end{equation}
This determines $\mathbf{\hat k}$.  Note that $\mathbf{\hat kk}$ is a
projector that removes $\mathbb{B}$.

Using this in Eq.~(\ref{BiaxialStiffness}), one can write:
\begin{equation}
\mathbf{v} = \frac{dq}{dt}\mathbf{\hat kL}_x,
\end{equation}
which can be integrated:
\begin{equation}
\mathbf{x} = \mathbf{x}_0 + q(t)\mathbf{\hat kL}_x.
\label{eq:positions}
\end{equation}
Here $\mathbf{x}$ is a vector
containing the positions of the particles and walls.
Eq.~(\ref{eq:positions} shows
that the position of each particle moves back and forth
on a line defined by the appropriate components of $\mathbf{\hat kL}_x$. 
So no area is traced out by position cycles, and ratcheting does not occur.

But if there are sliding contacts, then $\mathbf{k}$ does not remain
constant.  Recall that $\mathbf{k}$ changes whenever a contact
changes status.  Thus when a contact starts or stops sliding, the
relation between $\mathbf{v}$ and $\mathbf{f}_\mathrm{ext}$ changes,
and the particle motion changes direction.  This is what we saw in
Fig.~\ref{NScontacts}.

Another way to get paths that are not lines is add another term to 
the left hand side of Eq.~(\ref{BiaxialStiffness}).  When we use
the forcing given in Eq.~(\ref{2Dforcing}) and (\ref{2Dqs}), then
Eq.~(\ref{BiaxialStiffness}) becomes
\begin{equation}
		\frac{dq_x}{dt}\mathbf{L}_x  + \frac{dq_y}{dt}\mathbf{L}_y 
= \mathbf{kv},
		\label{2DBiaxialStiffness}
\end{equation}
and thus the motion is
\begin{equation}
\mathbf{x} = \mathbf{x}_0 + q_x(t) \mathbf{\hat kL}_x
	+ q_y(t) \mathbf{\hat kL}_y.
\end{equation}
Thus if the forcing traces out an area in the $(q_x,q_y)$ plane, than
each contact traces out a proportional area in the relative position plane.
This, together with Eq.~\ref{eq:area},
explains the result in Fig.~\ref{fig:ellipse}.

\end{document}